\documentclass[11pt]{article}
\usepackage[cp1251]{inputenc}
\usepackage[english]{babel}
\usepackage{amssymb,amsmath}
\usepackage{graphicx}
\usepackage{bm}
\begin{document}

\begin{center}
{\bf SU(2s+1) symmetry and nonlinear dynamic equations of spin s magnets}
\bigskip

{\bf M.Y. Kovalevsky, A.V. Glushchenko}
\medskip

{\small {\sf Kharkov Institute of Physics and Technology,   \\
     Academicheskaya 1, Kharkov, 61108, Ukraine,\\
     e-mail: $\underline{\mbox{mikov51@mail.ru}}$}}
\end{center}

\begin{abstract}
The article is devoted to the description of dynamics of magnets with arbitrary spin on the basis of the Hamiltonian formalism. The relationship between the magnetic ordering and Poisson bracket subalgebras of the magnetic degrees of freedom for spin s=1/2; 1; 3/2 has been established. We have been obtained non-linear dynamic equations without damping for normal and degenerate non-equilibrium states of high-spin magnets with the properties of the SO(3), SU(4), SU(2)$\times$SU(2), SU(3), SO(4), SO(5) symmetry of exchange interaction. The connection between models of the magnetic exchange energy and the Casimir invariants has been discussed.
\end{abstract}

keywords: hamiltonian mechanics, spin, symmetry, dynamics, Poisson brackets.

\section{Introduction}	
Currently, the studies of magnets, particles which have a spin  $s\geq1$, arouse much interest. These studies are pressing because of theoretical and experimental works of quasi-crystalline structures created on the optical lattices technology [1,2]. The capability to control geometrical parameters of the lattice and the intensity of the interparticle interaction makes them attractive in the study of magnetic properties of quantum objects. The additional incentive is associated with the Bose-Einstein condensation of neutral atoms with a non-zero spin [3,4].

With the increase of the particle spin value, the set of values required for a macroscopically complete description of non-equilibrium states is expanded. In contrast to the well-known Landau-Lifshitz equation [5], which describes the evolution of medium only with the help of the spin vector and is well-proven for the spin s=1/2. More complex magnetic systems with the spin  $s\geq1$ require the introduction of new dynamic values. The properties of symmetry of the magnetic exchange interaction with the spin  $s\geq1$, leads to a more complex structure of equilibrium states and non-equilibrium dynamic processes. These magnets in degenerate states have several types of breaking symmetry due to the transformation properties of the order parameters, which have a vector or tensor nature. Non-equilibrium processes of magnets with the spin s=1 have been analyzed in [6-11].

The structure of the paper is as follows: in section I, we have analyzed on the basis of the Hamiltonian approach the magnetic degrees of freedom, which characterize the macroscopic state of high-spin magnets, for which the Poisson bracket algebra has been obtained. In section II, we have discussed the symmetry properties of the Hamiltonian and formulated the differential conservation laws. We have obtained flux densities of dynamic values for normal and degenerate states. The relationship of Casimir invariants and models of the exchange magnetic energy has been discussed. In sections III-V, we have obtained non-linear dynamic equations for normal and degenerate states of magnets with the spin s=1/2; 1; 3/2 without damping.

\section{Variation principle and Poisson brackets in magnets}	
According to the general approach of continuum mechanics, let us present the Lagrangian of an arbitrary physical system as:  $L=L_{k}-H=\int{d^{3}xF_{a}({\bf x}, \phi({\bf x}'))\dot{\phi}_{a}({\bf x})}-H$, where  $L_{k}(\phi, \dot{\phi})$ is the kinematic part of the Lagrangian,  $H(\phi)=\int{d^3xe({\bf x},\phi)}$ is the Hamiltonian of the system. The energy density of the medium $e({\bf x},\phi({\bf x}'))$  and variable  $F_{a}({\bf x},\phi({\bf x}'))$ are the defined functionals of dynamic variables  $\phi_{a}(\bf x)$. The principle of stationary action leads to dynamic equations for the variables $\phi_{a}(\bf x)$:
\begin{equation}	
\label{eq1}
\dot{\phi}_{a}({\bf x})=\int{d^3x'J^{-1}_{ab}({\bf x, x}';\phi)\frac{\delta H(\phi)}{\delta\phi_b({\bf x}')}=\{\phi_{a}({\bf x}),H(\phi)}\}.
\end{equation}
The matrix  $J_{ab}({\bf x, x}';\phi)$, included in this formula, defined by the relation
\begin{align*}
J_{ab}({\bf x, x}';\phi)\equiv \frac{\delta F_{b}({\bf x}', \phi)}{\delta\phi_{a}({\bf x})}-\frac{\delta F_{a}({\bf x}, \phi)}{\delta\phi_{b}({\bf x}')}.
\end{align*}
This matrix defines the Poisson brackets for the variables $\phi_{a}(\bf x)$ as follows:
\begin{align*}
\{\phi_{a}({\bf x}),\phi_{b}({\bf x}')\}=J^{-1}_{ab}({\bf x, x}';\phi).
\end{align*}
The finite transformations  $\phi_{a}({\bf x})\rightarrow\phi_{a}'({\bf x})=\phi_{a}({\bf x}, \phi({\bf x}'))$, leaving the kinematic part of Lagrangian invariant
\begin{align*}
L_{k}(\phi, \dot{\phi})=\int{d^{3}xF_{a}({\bf x}; \phi)\dot{\phi}_{a}({\bf x})}=L_{k}(\phi',\dot{\phi}')=\int{d^{3}x F_a({\bf x};\phi')\dot{\phi}'_a({\bf x})}
\end{align*}
are canonical if the following formula is true: $F_b({\bf x}'; \phi)=\int{d^3xF_a({\bf x};\phi')}\delta\phi'_a({\bf x})/\delta\phi_b({\bf x}')$. In the case of infinitesimal transformations $\phi_{a}({\bf x})\rightarrow\phi_{a}'({\bf x})=\phi_{a}({\bf x})+\delta\phi_{a}({\bf x}, \phi({\bf x}'))$ we have
\begin{equation}	
\label{eq2}
\delta\phi_{a}({\bf x})=\{\phi_{a}({\bf x}), G\},~~~ G(\phi)=\int{d^3xF_a({\bf x},\phi)\delta\phi_a({\bf x})},
\end{equation}
where $G(\phi)$ is the generator of infinitesimal canonical transformations.

The formulation of the Hamiltonian approach to the description of non-equilibrium processes in magnets requires a choice of certain sets of dynamic variables, for which the Poisson structure is introduced. In the studied magnets, we start with the expression for the kinematic part of the Lagrangian in the form
\begin{equation}	
\label{eq3}
L_k({\bf x})=b_{\alpha\beta}({\bf x})\dot{a}_{\beta\alpha}({\bf x})\equiv tr\hat{b}({\bf x})\dot{\hat{a}}({\bf x}).
\end{equation}
Here $b_{\alpha\beta}$ and $a_{\alpha\beta}$ are the Hermitian $(2s+1)\times(2s+1)$ matrices $(\hat{a}=\hat{a}^+, \hat{b}=\hat{b}^+)$. The spin of particle variable $s$  can possess integer or half-integer values. For convenience and brevity, we will omit the spin indices of the matrix elements where it is possible.

Let us find the infinitesimal canonical transformations $\delta\phi_a({\bf x};\phi)$, leaving the kinematic part of the Lagrangian invariant. Knowing their explicit form, on the one hand, and taking into account the formula {\upshape\eqref{eq2}}, on the other hand, it is not difficult to find the Poisson brackets of the dynamic variables. Variations  $\delta b_{\alpha\beta}({\bf x})=0, \delta a_{\alpha\beta}({\bf x})\neq0$, (functions $\delta a_{\alpha\beta}({\bf x})$ do not depend on $\hat{a}({\bf x}), \hat{b}({\bf x})$), leave the kinematic part of the Lagrangian invariant, and according to {\upshape\eqref{eq2}}  can be represented as follows
\begin{equation}	
\label{eq4}
\delta a_{\alpha\beta}({\bf x})=\{a_{\alpha\beta}({\bf x}), G\}, ~~~~ \delta b_{\alpha\beta}({\bf x})=\{b_{\alpha\beta}({\bf x}), G\},
\end{equation}
here $G=\int{d^3xb_{\alpha\beta}({\bf x})\delta a_{\beta\alpha}({\bf x})}$ is the generator of these transformations. Comparing {\upshape\eqref{eq2}} and {\upshape\eqref{eq4}}, we will obtain the Poisson brackets
\begin{equation}	
\label{eq5}
\{b_{\alpha\beta}({\bf x}), b_{\mu\nu}({\bf x'})\}=0, ~~~~ \{b_{\alpha\beta}({\bf x}), a_{\mu\nu}({\bf x'})\}=-\delta_{\alpha\nu}\delta_{\beta\mu}\delta({\bf x-x'}).
\end{equation}
For the kinematic part of the Lagrangian $L_{k}({\bf x})=-tr\dot{\hat{b}}({\bf x})\hat{a}({\bf x})$,which differs from {\upshape\eqref{eq3} by the full-time derivative, we will obtain the following in a similar way:
\begin{equation}
\label{eq6}
\{a_{\alpha\beta}({\bf x}), a_{\mu\nu}({\bf x'})\}=0, ~~~~ \{a_{\alpha\beta}({\bf x}), b_{\mu\nu}({\bf x'})\}=\delta_{\alpha\nu}\delta_{\beta\mu}\delta({\bf x-x'}).
\end{equation}
Matrices $\hat{a}$  and $\hat{b}$ are canonically conjugate values. Let us connect these matrices with the physical values of the theory of magnetism. For this purpose, let us introduce the matrix
\begin{equation}
\label{eq7}
\hat{g}({\bf x})\equiv i\left[\hat{b}({\bf x}),\hat{a}({\bf x})\right].
\end{equation}
The square brackets here and below denote the commutator of two matrices. These matrices make sense of the density of  $SU(2s+1)$ symmetry generators. The conditions of tracelessness and hermiticity of this matrix leads to a number of real parameters $4s(2s+1)$, which are generally not independent. Using the form {\upshape\eqref{eq7}} and formulas {\upshape\eqref{eq5}}, {\upshape\eqref{eq6}}, we obtain the Poisson bracket for the generator density of the SU(2s+1) symmetry:
\begin{equation}
\label{eq8}
i\{g_{\alpha\beta}({\bf x}), g_{\gamma\rho}({\bf x}')\}=(g_{\gamma\beta}({\bf x})\delta_{\alpha\rho}-g_{\alpha\rho}({\bf x})\delta_{\gamma\beta})\delta({\bf x-x'}).
\end{equation}
Here $\delta_{\alpha\beta}$ - Kronecker symbol. Casimir invariants of this Poisson bracket algebra are as follows
\begin{equation}
\label{eq9}
g_n({\bf x})\equiv tr\hat{g}^n({\bf x}), ~~~~ \{g_n({\bf x}), \hat{g}({\bf x}')\}=0,
\end{equation}
where $n=2,3,..,2s+1$. The presence of 2s Casimir invariants leads to the number of independent magnetic degrees of freedom 2s(2s+1). With these variables, we describe normal states of the studied magnets. The symmetry of normal equilibrium states and the Hamiltonian symmetry are the same. This ordering case describes paramagnetic states. The Hamiltonian for these states is a functional of the matrix  {\upshape\eqref{eq7}}:  $H=H(\hat{g}({\bf x}))$. For normal states, the exchange energy density is a function of the matrix $\hat{g}({\bf x})$ and its gradient  $e({\bf x})=e(\hat{g}({\bf x}),\nabla\hat{g}({\bf x}))$. Using {\upshape\eqref{eq1}}, {\upshape\eqref{eq8}}, we get the dynamic equation of high-spin magnets for the matrix  $\hat{g}({\bf x})$
\begin{equation}
\label{eq10}
\dot{\hat{g}}({\bf x})=i\left[\hat{g}({\bf x}), \frac{\delta\hat{H}(\hat{g})}{\delta g({\bf x})}\right],
\end{equation}
which generalizes the Landau-Lifshitz equation for the case of magnets with arbitrary spin  s. The influence of the spin manifested in the rank of this matrix and in the number of its independent matrix elements.

Let us introduce symmetric and antisymmetric parts for the matrix  $\hat{g}$: $\hat{g}\equiv \hat{g}^{(s)}+i\hat{g}^{(a)}$, where
\begin{align*}
g_{\alpha\beta}^{(s)}=\frac{1}{2}(g_{\alpha\beta}+g_{\beta\alpha}), ~~~~ g_{\alpha\beta}^{(a)}=\frac{1}{2i}(g_{\alpha\beta}-g_{\beta\alpha}).
\end{align*}
These matrices are real. It is easy to find the Poisson brackets for them
\begin{align*}
\{g_{\alpha\beta}^{(a)}({\bf x}), g_{\gamma\rho}^{(a)}({\bf x'})\}=\delta({\bf x-x'})(\delta_{\alpha\rho}g_{\beta\gamma}^{(a)}({\bf x})+\delta_{\beta\gamma}g_{\alpha\rho}^{(a)}({\bf x})+\delta_{\alpha\gamma}g_{\rho\beta}^{(a)}({\bf x})+\delta_{\beta\rho}g_{\gamma\alpha}^{(a)}({\bf x}))/2, \\
\{g_{\alpha\beta}^{(s)}({\bf x}), g_{\gamma\rho}^{(s)}({\bf x}')\}=-\delta({\bf x-x'})(\delta_{\alpha\rho}g_{\beta\gamma}^{(a)}({\bf x})+\delta_{\beta\gamma}g_{\alpha\rho}^{(a)}({\bf x})+\delta_{\alpha\gamma}g_{\beta\rho}^{(a)}({\bf x})+\delta_{\beta\rho}g_{\alpha\gamma}^{(a)}({\bf x}))/2, \\
\{g_{\alpha\beta}^{(a)}({\bf x}), g_{\gamma\rho}^{(s)}({\bf x}')\}=\delta({\bf x-x'})(-\delta_{\alpha\rho}g_{\beta\gamma}^{(s)}({\bf x})+\delta_{\beta\gamma}g_{\alpha\rho}^{(s)}({\bf x})-\delta_{\alpha\gamma}g_{\beta\rho}^{(s)}({\bf x})+\delta_{\beta\rho}g_{\gamma\alpha}^{(s)}({\bf x}))/2.
\end{align*}
As we can see, the Poisson bracket algebra of antisymmetric matrices is closed. If the magnetic Hamiltonian $H=H(\hat{g}^{(s)})$ depends only on these matrices, we get the dynamic equation
\begin{equation}
\label{eq11}
\dot{\hat{g}}^{(a)}({\bf x})=\left[\hat{g}^{(a)}({\bf x}), \frac{\delta\hat{H}(\hat{g}^{(a)})}{\delta g^{(a)}({\bf x})}\right].
\end{equation}
This equation is a meaningful for magnets with the spin  $s\geq1$.

Another important class of magnetic states is the degenerate states. The symmetry of these equilibrium states less than the Hamiltonian symmetry. In this case, the magnetic degrees of freedom consist of the matrices  $\hat{g}({\bf x})$ and $\hat{a}({\bf x})$. The latter is the order parameter for degenerate states of high-spin magnets. Formulas {\upshape\eqref{eq5}}-{\upshape\eqref{eq7}} allow us to obtain the Poisson brackets of the matrices   $\hat{g}({\bf x})$ and $\hat{a}({\bf x})$
\begin{equation}
\label{eq12}
i\{a_{\alpha\beta}({\bf x}), g_{\gamma\rho}({\bf x}')\}=(a_{\gamma\beta}({\bf x})\delta_{\alpha\rho}-a_{\alpha\rho}({\bf x})\delta_{\gamma\beta})\delta({\bf x-x'}).
\end{equation}

Let us note that due to {\upshape\eqref{eq12}},  $\{tr\hat{a}({\bf x}), g_{\gamma\rho}({\bf x}')\}=0$, therefore, in view of the linearity of the right-hand side of  {\upshape\eqref{eq12}}, we can assume that   $tr\hat{a}=0$. It is easy to prove that the extended algebra of the Poisson brackets {\upshape\eqref{eq8}}, {\upshape\eqref{eq12}} has the Casimir invariants
\begin{equation}
\label{eq13}
a_n({\bf x})\equiv tr\hat{a}^n({\bf x}), ~~~~ \{a_n({\bf x}), \hat{g}({\bf x}')\}=0, ~~~~ \{a_n({\bf x}), \hat{a}({\bf x}')\}=0.
\end{equation}
The Hamiltonian of states with spontaneously broken symmetry is a functional of the two matrices $H=H(\hat{g}({\bf x}), \hat{a}({\bf x}))$. For degenerate states, the energy density  $e=e(\hat{g}({\bf x}), \nabla\hat{g}({\bf x}), \hat{a}({\bf x}), \nabla\hat{a}({\bf x}))$ is the function of both matrices and their gradients. In this case, due to {\upshape\eqref{eq1}}, {\upshape\eqref{eq8}}, {\upshape\eqref{eq12}}, the dynamic equations for high-spin magnets take the form
\begin{align*}
\dot{\hat{g}}({\bf x})=i\left[\hat{g}({\bf x}), \frac{\delta\hat{H}(\hat{g}, \hat{a})}{\delta g({\bf x})}\right]+i\left[\hat{a}({\bf x}), \frac{\delta\hat{H}(\hat{g}, \hat{a})}{\delta a({\bf x})}\right], ~~~~ \dot{\hat{a}}({\bf x})=i\left[\hat{a}({\bf x}), \frac{\delta\hat{H}(\hat{g}, \hat{a})}{\delta g({\bf x})}\right].
\end{align*}
\section{Conservation laws. Hamiltonian symmetry and models}	
Main interactions in magnetic systems have an exchange nature. The Hamiltonian  $H=\mathcal{H}+V\equiv\int{d^3xe({\bf x})}$ includes strong exchange interaction and weak and less symmetrical relativistic interaction  $V$. The symmetry of the exchange Hamiltonian and equilibrium state allow us to establish a set of thermodynamic parameters describing macroscopic magnetic states. For simplicity, we will not further consider weak magnetic interactions, $H\sim\mathcal{H}$.

The analysis of dynamic processes in continuous media requires the formulation of conservation laws in the differential form, taking into account the Hamiltonian symmetry. The  SU(2)$\sim$SO(3) symmetry of the exchange Hamiltonian   $\{S_{\alpha}, H\}=0$ corresponds to magnets with the spin $s=1/2$. The set of integrals of motion  $\gamma_a$ consists of the Hamiltonian and the spin moment  $\gamma_a=H, S_{\alpha}=\int{d^3x\zeta_a({\bf x})}$. Here variables  $\zeta_a({\bf x})=e({\bf x}), s_{\alpha}({\bf x})$ are the density of the additive integrals of motion  ($a=0, \alpha$). Using the representation of flux densities of additive integrals of motion [12], we will obtain the dynamic equations reflecting the conservation laws in the differential form: for the energy density
\begin{equation}
\label{eq31}
\dot{e}({\bf x})=-\nabla_kq_k({\bf x}), ~~~~ q_k({\bf x})=\frac{1}{2}\int{d^3x'x'_k}\int_0^1{d\lambda\{e({\bf x}+\lambda{\bf x}'), e({\bf x}-(1-\lambda){\bf x}')\}},
\end{equation}
where $q_k({\bf x})$ is the flux energy density. For the density of generator SU(2) symmetry we have
\begin{equation}
\label{eq32}
\dot{s}_{\alpha}({\bf x})=-\nabla_kj_{\alpha k}({\bf x}), ~~~~ j_{\alpha k}({\bf x})=\int{d^3x'x'_k}\int_0^1{d\lambda\{s_{\alpha}({\bf x}+\lambda{\bf x}'), e({\bf x}-(1-\lambda){\bf x}')\}},
\end{equation}
where $j_{\alpha k}({\bf x})$ is the flux spin density. While obtaining the last-mentioned equation, the symmetry of the exchange energy density has been considered
\begin{equation}
\label{eq33}
\{S_{\alpha}, e({\bf x})\}=0.
\end{equation}

In case of the arbitrary spin $s$ and the SU(2s+1) symmetry of the Hamiltonian   $\{G_{\alpha\beta}, H\}=0$, the set of integrals of motion consists of the Hamiltonian and the matrix  $G_{\alpha\beta}$ of the rank 2s+1:  $\gamma_{a}=H, G_{\alpha\beta}=\int{d^3x\zeta_a({\bf x})}$, ($a=0, \alpha\beta$). To establish the dynamic equations of densities of additive integrals of motion $\zeta_a({\bf x})=e({\bf x}), g_{\alpha\beta}({\bf x})$, the symmetry of the exchange energy density will be considered:
\begin{equation}
\label{eq34}
\{G_{\alpha\beta}, e({\bf x})\}=0.
\end{equation}
Using this formula and the representation of flux densities [12], we get the differential conservation law
\begin{equation}
\label{eq35}
\dot{\hat{g}}({\bf x})=-\nabla_k\hat{j}_k({\bf x}), ~~~~ \hat{j}_k({\bf x})=\int{d^3x'x'_k}\int_0^1{d\lambda\{\hat{g}({\bf x}+\lambda{\bf x}'), e({\bf x}-(1-\lambda){\bf x}')\}}.
\end{equation}

Here $\hat{j}_k({\bf x})$ is the flux density corresponding to the conserved variable  $\hat{G}$. The motion equation for the energy density in the differential form and the expression for its flux density in  {\upshape\eqref{eq31}} will not change. In addition to the properties of the symmetry {\upshape\eqref{eq33}} or {\upshape\eqref{eq34}}, the density of the exchange energy is translation invariant and invariant with respect to rotations in the configuration space
\begin{align*}
\{P_k, e({\bf x})\}=\nabla_ke({\bf x}), ~~~~ \{L_i, e({\bf x})\}=\varepsilon_{ikl}x_k\nabla_le({\bf x}).
\end{align*}
The momentum and angular momentum of the magnetic system in terms of canonical variables are defined by formulas $P_k=\int{d^3x\pi_k({\bf x})}$, $L_k=\varepsilon_{kij}\int{d^3xx_i\pi_j({\bf x})}$. Here $\pi_k({\bf x})\equiv-tr\hat{b}({\bf x})\nabla_k\hat{a}({\bf x})$ is the density of the magnon momentum.

In case of normal non-equilibrium states, basing on formulas {\upshape\eqref{eq31}}, {\upshape\eqref{eq35}}, {\upshape\eqref{eq8}}, it is easy to express flux densities of additive integral of motion in terms of the Hamiltonian
\begin{equation}
\label{eq36}
\hat{j}_k=i\left[\hat{g}, \frac{\partial\hat{e}(\hat{g})}{\partial\nabla_kg}\right], ~~~~ q_k=tr\frac{\delta\hat{H}(\hat{g})}{\delta g}\hat{j}_k.
\end{equation}

For degenerate states, we will obtain flux densities of additive integrals of motion in a similar way
\begin{equation}
\label{eq37}
\hat{j}_k=i\left[\hat{g}, \frac{\partial\hat{e}(\hat{g}, \hat{a})}{\partial\nabla_kg}\right]+i\left[\hat{a}, \frac{\partial\hat{e}(\hat{g}, \hat{a})}{\partial\nabla_ka}\right], ~~~~
q_k=itr\frac{\delta\hat{H}(\hat{g}, \hat{a})}{\delta a}\left[\hat{a}, \frac{\partial\hat{e}(\hat{g}, \hat{a})}{\partial\nabla_kg}\right]+tr\frac{\delta\hat{H}(\hat{g}, \hat{a})}{\delta g}\hat{j}_k.
\end{equation}

The dynamic equations  {\upshape\eqref{eq31}}, {\upshape\eqref{eq35}} and {\upshape\eqref{eq36}} describe non-equilibrium processes for the normal states of high-spin magnets in the adiabatic approximation with the SU(2s+1) Hamiltonian symmetry. Similar dynamics of degenerate non-equilibrium states with the same Hamiltonian symmetry will be defined by the equations {\upshape\eqref{eq31}}, {\upshape\eqref{eq35}}, {\upshape\eqref{eq37}}.

Let us consider normal non-equilibrium states with the SU(2s +1) symmetry. The dependence of the exchange Hamiltonian on the Hermitian matrix will be modeled as follows
\begin{equation}
\label{eq38}
e=e_o+e_n, ~~~~  e_o=J tr \hat{g}^2/2, ~~~~  e_n=\bar{J}tr(\nabla\hat{g})^2/2>0,
\end{equation}
here $e_o$ is the homogeneous part of the exchange energy, which we choose as the function of the Casimir invariant $g_2$, $J$ is the constant of homogeneous exchange interaction. For simplicity, we will not consider the effect of other Casimir invariants with $n\geq3$. The term   $e_n$ is the contribution of the inhomogeneous exchange energy, and it is essential in the description of the evolution of the magnets, $\bar{J}$  is the constant of inhomogeneous exchange interaction. The exchange interaction, in combination with the relativistic interactions, will determine the equilibrium values of the matrix  $\hat{g}$ from the condition $\delta\hat{H}/\delta g=0$. Considering formulas {\upshape\eqref{eq36}}, {\upshape\eqref{eq38}}, we will obtain the equation
\begin{align*}
\dot{\hat{g}}=-i\bar{J}\left[\hat{g}, \triangle\hat{g}\right].
\end{align*}
This equation is the SU(2s+1) symmetric generalization of the non-linear equation {\upshape\eqref{eq45}}, true for magnets with the spin $s=1/2$.

Let us consider degenerate states. In this case, we will choose the energy density as the function of Casimir invariants {\upshape\eqref{eq9}} for the algebra  {\upshape\eqref{eq8}} corresponding to normal states, and the Casimir invariants {\upshape\eqref{eq13}} of the extended algebra  {\upshape\eqref{eq8}},  {\upshape\eqref{eq12}}. For simplicity, only Casimir invariants quadratic by the matrices $\hat{g}, \hat{a}$ will be used:
\begin{equation}
\label{eq39}
e_o=Jtr\hat{g}^2/2+Atr\hat{a}^2/2, ~~~~ e_n=\bar{J}tr(\nabla\hat{g})^2/2+Btr(\nabla\hat{a})^2/2.
\end{equation}
Taking into account  {\upshape\eqref{eq39}}, non-linear dynamic equations of degenerate states take the form
\begin{align*}
\dot{\hat{g}}=-i\bar{J}\left[\hat{g}, \triangle\hat{g}\right]-iB\left[\hat{a}, \triangle\hat{a}\right], ~~~~ \dot{\hat{a}}=i\left[\hat{a},J\hat{g}-\bar{J}\triangle\hat{g}\right].
\end{align*}
If the spin value is large enough and the  SU(2s+1) Hamiltonian symmetry is high, the resulting non-linear matrix equations are generally difficult to analyze. In the following paragraphs, we will consider the spin values $s=1/2;1;3/2$, study the Poisson bracket subalgebras and derive the non-linear dynamic equations of high-spin magnets in simpler cases.
\section{Magnets with the spin s=1/2}
For the magnets with the spin  s=1/2, the matrix $\hat{g}$ has the dimensions of $2\times2$ and it is convenient to depict it as an expansion for irreducible Pauli matrices  $\hat{\sigma}_{\alpha}$
\begin{equation}
\label{eq41}
\hat{g}({\bf x})=s_{\alpha}({\bf x})\hat{\sigma}_{\alpha}\equiv \begin{pmatrix} s_{z}({\bf x}) & s_x({\bf x})-is_y({\bf x}) \\ s_x({\bf x})+is_y({\bf x}) & -s_z({\bf x}) \end{pmatrix},
\end{equation}
where $s_{\alpha}({\bf x})=tr\hat{\rho}({\bf x})\hat{\sigma}_{\alpha}/2$ is the average spin moment. The polarization density matrix  $\hat{\rho}({\bf x})\equiv\hat{I}/2+\hat{g}({\bf x})$  describes the quantum states of magnets in the single-particle approximation,  $\hat{I}$ is the unit $2\times2$ matrix. The constraint $tr\hat{\rho}^2\leq1$ defines the spin vector variation range. For pure state,  $tr\hat{\rho}^2=tr\hat{\rho}=1$, from which $s=1/2$,the spin vector direction is arbitrary. For mixed states,  $tr\hat{\rho}^2<tr\hat{\rho}$, therefore  $0\leq s\leq1/2$. For magnets with spin s = 1/2 the spin vector completely defines the magnetic degrees of freedom of the media. By virtue of {\upshape\eqref{eq8}}, {\upshape\eqref{eq41}}, we will find the Poisson bracket for the spin vector
\begin{equation}
\label{eq42}
\{s_{\alpha}({\bf x}),s_{\beta}({\bf x}')\}=\delta({\bf x-x'})\varepsilon_{\alpha\beta\gamma}s_{\gamma}({\bf x}).
\end{equation}

$\bf Case$ $\bf1$. Using the functional hypothesis  $H=H(s({\bf x}))$, we obtain the Landau-Lifshitz dynamic equation [5]:
\begin{align*}
\dot{s}_{\alpha}({\bf x})=\varepsilon_{\alpha\beta\gamma}\frac{\delta H({\bf s})}{\delta s_{\beta}({\bf x})}s_{\gamma}({\bf x}).
\end{align*}
If the SO(3) symmetry property {\upshape\eqref{eq33}} is true, then {\upshape\eqref{eq32}} leads to the equation
\begin{equation}
\label{eq43}
\dot{s}_{\alpha}=-\nabla_kj_{\alpha k}, ~~~~ j_{\alpha k}=\varepsilon_{\alpha\beta\gamma}\frac{\partial e}{\partial\nabla_ks_{\beta}}s_{\gamma}.
\end{equation}
The further specification of the dynamic equation is related to the choice of the Hamiltonian model as a spin vector functional. For magnets with the spin s=1/2, the Heisenberg Hamiltonian is commonly used
\begin{center}
$H=\int{d^3xe({\bf x})}=-\int{d^3xd^3x'J({|\bf x-x'}|)s_{\alpha}({\bf x})s_{\alpha}({\bf x}')},$
\end{center}
where $J({|\bf x-x'}|)$ is the exchange integral of two-particle magnetic interaction. With accuracy of squared terms of spatial gradients of the spin density, we will imagine the expression of magnetic energy density corresponding to this Hamiltonian
\begin{equation}
\label{eq44}
e({\bf x})=-Js_{\alpha}({\bf x})s_{\alpha}({\bf x})+\frac{1}{2}\bar{J}\nabla_ks_{\alpha}({\bf x})\nabla_ks_{\alpha}({\bf x}),
\end{equation}
where $J\equiv\int{d^3xJ(|\bf x|)}$, $\bar{J}\equiv\int{d^3xx^2J(|{\bf x}|)/3>0}$ are effective exchange integrals of two-particle interaction. The first and second terms in  {\upshape\eqref{eq44}} describe accordingly homogeneous and inhomogeneous exchange interaction. The functional form of the homogeneous part of the energy is defined by the Casimir invariant $s^2_{\alpha}$  of the  algebra {\upshape\eqref{eq42}}. The equation {\upshape\eqref{eq43}} with regard to {\upshape\eqref{eq44}} will take the form
\begin{equation}
\label{eq45}
\dot{s}_{\alpha}({\bf x})=-\bar{J}\varepsilon_{\alpha\beta\gamma}\triangle s_{\beta}({\bf x})s_{\gamma}({\bf x}).
\end{equation}
The analysis of non-linear solutions of this equation in case spatial dimension $d=1$  has been previously considered, in particular, in the papers [13-17]. For higher spatial dimensions $d=2,3$  solutions of this equation and their analysis were carried out in [17-22].

Let us consider magnets with spontaneously broken SO(3) symmetry. This symmetry breaking can be uniaxial and biaxial.

$\bf Case$ $\bf2$: The state with the uniaxial spontaneous symmetry breaking. Specific magnets, where such ordering is implemented, are antiferromagnets [19,23] and A-phase of superfluid $He^3$[24]. These states will be further characterized by the spin vector and antiferromagnetic vector  $n_{\alpha}({\bf x})$, which can be expressed in terms of the order parameter matrix  $\hat{a}({\bf x})$. The Poisson bracket algebra consists of formulas {\upshape\eqref{eq42}} and
\begin{equation}
\label{eq46}
\{s_{\alpha}({\bf x}), n_{\beta}({\bf x}')\}=\delta({\bf x-x'})\varepsilon_{\alpha\beta\gamma}n_{\gamma}({\bf x}), ~~~~ \{n_{\alpha}({\bf x}), n_{\beta}({\bf x}')\}=0.
\end{equation}
It contains two Casimir invariants:  $\bf sn$ and $n^2$. It is generally assumed that  $n^2=1$. The energy density $e=e(\bf s, \nabla s, n, \nabla n)$ is a functional of the spin density and the unit vector of the spin anisotropy. According to formulas {\upshape\eqref{eq42}}, {\upshape\eqref{eq46}}, we obtain the dynamic equations
\begin{align*}
\dot{s}_{\alpha}=\varepsilon_{\alpha\beta\gamma}\left(\frac{\delta H}{\delta s_{\beta}}s_{\gamma}+\frac{\delta H}{\delta n_{\beta}}n_{\gamma}\right), ~~~~ \dot{n}_{\alpha}=\varepsilon_{\alpha\beta\gamma}\frac{\delta H}{\delta s_{\beta}}n_{\gamma}.
\end{align*}
The first of these equations, taking into account SO(3) symmetry property  {\upshape\eqref{eq32}}, take the form
\begin{align*}
\dot{s}_{\alpha}=-\nabla_kj_{\alpha k}, ~~~~ j_{\alpha k}=\varepsilon_{\alpha\beta\gamma}\left(\frac{\partial e}{\partial\nabla_ks_{\beta}}s_{\gamma}+\frac{\partial e}{\partial\nabla_kn_{\beta}}n_{\gamma}\right).
\end{align*}
Microscopic theory, which gives the expression of the exchange magnetic energy of antiferromagnetic even for systems with spin $s=1/2$is so far lacking. Therefore, hereinafter we use the model representations of the magnetic exchange energy. We represent this value as the sum of two terms:  $e=e_o+e_n$. The first of these is the density of the homogeneous part of the exchange energy. This value is the function of the Casimir invariants $({\bf sn}), s^2$,  and has the SO(3) symmetry: $e_o=Js^2/2+A({\bf sn})^2/2, J, A$  are the homogeneous exchange constants. The inhomogeneous exchange energy depicted as $e_n=Btr(\nabla_k{\bf n})^2/2$, here $B>0$ is an inhomogeneous exchange constant. As a result, non-linear dynamic equations take form
\begin{center}
$\dot{s}_{\alpha}=-B\varepsilon_{\alpha\beta\gamma}\triangle n_{\beta} n_{\gamma}, ~~~~ \dot{n}=\varepsilon_{\alpha\beta\gamma}n_{\gamma}(Js_{\beta}-B\triangle n_{\beta}).$
\end{center}

$\bf Case$ $\bf3$: Let us consider the biaxial symmetry breaking, which is realized in spin glasses and superfluid B-phase of He$^3$  [24].  In these magnets, a complete spontaneous breaking of the SO(3) symmetry take place. A set of magnetic degrees of freedom consists of the spin density  $\bf s$ and the orthogonal rotation matrix $\hat{R}$. The Poisson bracket algebra includes the formula  {\upshape\eqref{eq42}} and the relations
\begin{equation}
\label{eq47}
\{s_{\alpha}({\bf x}), R_{\beta\lambda}({\bf x}')\}=\delta({\bf x-x'})\varepsilon_{\alpha\beta\gamma}R_{\gamma\lambda}({\bf x}), ~~~~ \{R_{\alpha\gamma}({\bf x}), R_{\beta\rho}({\bf x}')\}=0.
\end{equation}
Formulas {\upshape\eqref{eq42}}, {\upshape\eqref{eq47}} lead to the dynamic equations
\begin{align*}
\dot{s}_{\alpha}=\varepsilon_{\alpha\beta\gamma}\left(\frac{\delta H}{\delta s_{\beta}}s_{\gamma}+\frac{\delta H}{\delta R_{\lambda\beta}}R_{\lambda\gamma}\right), ~~~~ \dot{R}_{\alpha\beta}=R_{\alpha\rho}\varepsilon_{\rho\beta\gamma}\frac{\delta H}{\delta s_{\gamma}}.
\end{align*}
The SO(3) symmetry property of the energy density {\upshape\eqref{eq33}} allows us to find the equation of the spin density as
\begin{align*}
\dot{s}_{\alpha}=-\nabla_kj_{\alpha k}, ~~~~ j_{\alpha k}=\varepsilon_{\alpha\beta\gamma}\left(\frac{\partial e}{\partial\nabla_ks_{\beta}}s_{\gamma}+\frac{\partial e}{\partial\nabla_kR_{\lambda\beta}}R_{\lambda\gamma}\right).
\end{align*}
Let us express the exchange energy density  $e=e_o+e_n$ as $e_o=Js^2/2+Atr\hat{R}^2/2$, $e_n=Btr(\nabla\hat{R})^2/2$. As a result, the dynamic equations will take form
\begin{center}
$\dot{s}_{\alpha}=-B\varepsilon_{\alpha\beta\gamma}\triangle R_{\lambda\beta}R_{\lambda\gamma}, ~~~~ \dot{R}_{\alpha\beta}=-JR_{\alpha\rho}\varepsilon_{\rho\beta\gamma}s_{\gamma}.$
\end{center}
\section{Magnets with the spin s=1}
Let us consider the magnets, particles of which have the spin  s=1. The state of such magnets can be described using the polarization matrix $\hat{\rho}({\bf x})=\hat{I}/3+\hat{g}({\bf x})$ with the dimension of  3$\times$3, which include eight real values. We will expand the polarization matrix into a complete set of irreducible  (3$\times$3) matrices
\begin{center}
$\hat{\rho}({\bf x})=\hat{I}/3+s_{\alpha}({\bf x})\hat{s}_{\alpha}/2+q_{\alpha\beta}({\bf x})\hat{q}_{\beta\alpha}.$
\end{center}
The matrix elements of spin operators  $\hat{s}_{\alpha}$ and the quadrupole tensor $\hat{q}_{\beta\alpha}$ here will take the following form, according to [25]:
\begin{center}
$(\hat{s}_{\alpha})_{\mu\nu}\equiv-i\varepsilon_{\alpha\mu\nu}, ~~~~ (\hat{q}_{\beta\alpha})_{\mu\nu}\equiv(\delta_{\beta\mu}\delta_{\alpha\nu}+\delta_{\beta\nu}\delta_{\alpha\mu}-2\delta_{\beta\alpha}\delta_{\mu\nu}/3)/2,$
\end{center}
as well as satisfy the orthogonality conditions:
\begin{center}
$tr\hat{s}_{\alpha}=0, ~~~~ tr\hat{q}_{\alpha\beta}=0, ~~~~ tr\hat{s}_{\gamma}\hat{q}_{\alpha\beta}=0, ~~~~ tr\hat{s}_{\alpha}\hat{s}_{\beta}=2\delta_{\alpha\beta},$ \\$ tr\hat{q}_{\gamma\rho}\hat{q}_{\alpha\beta}=(\delta_{\alpha\gamma}\delta_{\beta\rho}+\delta_{\alpha\rho}\delta_{\beta\gamma}-2\delta_{\alpha\beta}\delta_{\gamma\rho}/3)/2, ~~~~ tr\hat{s}_{\alpha}\hat{s}_{\beta}\hat{s}_{\gamma}\hat{s}_{\rho}=\delta_{\alpha\beta}\delta_{\gamma\rho}+\delta_{\alpha\rho}\delta_{\gamma\beta}.$
\end{center}
Given these relations, we obtain the relation of average values of the spin density $s_{\alpha}({\bf x})$ and the quadrupole matrix $q_{\alpha\beta}({\bf x})$ in terms of the polarization matrix:
\begin{center}
$tr\hat{\rho}({\bf x})\hat{s}_{\alpha}=s_{\alpha}({\bf x}), ~~~~ tr\hat{\rho}({\bf x})\hat{q}_{\alpha\beta}=q_{\alpha\beta}({\bf x}).$
\end{center}
Therefore, we get
\begin{center}
$tr\hat{\rho}({\bf x})\hat{g}_{\alpha\beta}=g_{\alpha\beta}({\bf x})\equiv q_{\alpha\beta}({\bf x})-i\varepsilon_{\alpha\beta\gamma}s_{\gamma}({\bf x})/2,$
\end{center}
where $\hat{g}_{\alpha\beta}\equiv\hat{q}_{\alpha\beta}-i\varepsilon_{\alpha\beta\gamma}\hat{s}_{\gamma}/2$. The magnetic degrees of freedom of the considered matter consist of   the spin vector $s_{\alpha}({\bf x})$ and the quadrupole matrix  $q_{\alpha\beta}({\bf x})$. They are related to the matrix  $\hat{g}({\bf x})$ by the formulas
\begin{equation}
\label{eq51}
s_{\alpha}=i\varepsilon_{\alpha\beta\gamma}g_{\beta\gamma}, ~~~~ q_{\alpha\beta}=(g_{\alpha\beta}+g_{\beta\alpha})/2.
\end{equation}
It is easy to see that the Poisson brackets  {\upshape\eqref{eq42}} are true for the vectors $s_{\alpha}$ in view of {\upshape\eqref{eq8}}, {\upshape\eqref{eq51}}. For the remaining variables, we will find
\begin{equation}\label{eq52}
\begin{split}
&\{s_{\alpha}({\bf x}), q_{\beta\gamma}({\bf x}')\}=\delta({\bf x-x'})(\varepsilon_{\alpha\beta\rho}q_{\rho\gamma}({\bf x})+\varepsilon_{\alpha\gamma\rho}q_{\rho\beta}({\bf x})),  \\  &\{q_{\alpha\beta}({\bf x}), q_{\mu\nu}({\bf x}')\}=\delta({\bf x-x'})s_{\gamma}({\bf x})(\varepsilon_{\gamma\alpha\nu}\delta_{\beta\mu}+\varepsilon_{\gamma\beta\mu}\delta_{\alpha\nu}+\varepsilon_{\gamma\beta\nu}\delta_{\alpha\mu}+\varepsilon_{\gamma\alpha\mu}\delta_{\beta\nu})/4.
\end{split}
\end{equation}

Let’s consider the Hermitian matrix $\hat{a}({\bf x})$ and establish its connection with physical values by the formula
\begin{center}
$a_{\alpha\beta}({\bf x})\equiv w_{\alpha\beta}({\bf x})-i\varepsilon_{\alpha\beta\gamma}n_{\gamma}({\bf x})/2$.
\end{center}
We make sense the vector $\bf n$ the physical meaning of the antiferromagnetic vector. The tensor  $\hat{w}$ will have the meaning of the T-even order parameter, which is responsible for the nematic ordering. For the introduced values, we will obtain the following non-trivial Poisson brackets, according to  {\upshape\eqref{eq12}}
\begin{equation}\label{eq53}
\begin{split}
&\left\{s_{\alpha}({\bf x}), n_{\beta}({\bf x'})\right\}=\delta({\bf x-x'})\varepsilon_{\alpha\beta\gamma}n_{\gamma}({\bf x}), \\ &\left\{n_{\alpha}({\bf x}), q_{\beta\gamma}({\bf x'})\right\}=\delta({\bf x-x'})\left(\varepsilon_{\alpha\beta\rho}w_{\rho\gamma}({\bf x})+\varepsilon_{\alpha\gamma\rho}w_{\rho\beta}({\bf x})\right), \\ &\left\{s_{\alpha}({\bf x}), w_{\beta\gamma}({\bf x'})\right\}=\delta({\bf x-x'})\left(\varepsilon_{\alpha\gamma\rho}w_{\beta\rho}({\bf x})+\varepsilon_{\alpha\beta\rho}w_{\gamma\rho}({\bf x})\right), \\ &\left\{w_{\alpha\beta}({\bf x}), q_{\gamma\rho}({\bf x'})\right\}=\delta({\bf x-x'})n_{\gamma}({\bf x})(\varepsilon_{\alpha\nu\gamma}\delta_{\beta\mu}+\varepsilon_{\beta\mu\gamma}\delta_{\alpha\nu}+ \varepsilon_{\beta\nu\gamma}\delta_{\alpha\mu}+\varepsilon_{\alpha\mu\gamma}\delta_{\beta\nu})/4.
\end{split}
\end{equation}
Formulas {\upshape\eqref{eq42}}, {\upshape\eqref{eq52}}, {\upshape\eqref{eq53}} allow to identify Poisson bracket subalgebras and determine the dynamics of magnets with spin s=1 for all types of magnetic ordering. As it follows from these formulas the minimal Poisson bracket subalgebra will contain just the spin vector. This case is equivalent to the previously discussed case 1. Furthermore, this equation for magnets with the spin  s=1 is obtained from the equation {\upshape\eqref{eq11}}. The Poisson bracket subalgebra for spin and antiferromagnet vector lead to the dynamic equations, which are equivalent to the previously discussed case 2. It can be shown that the case of the complete SO(3) symmetry breaking, which is possible in magnets with the spin  s=1, coincides with the previously discussed case 3. Therefore, we will focus on new magnetic states, which were absent in magnets with the spin s=1/2. These include cases 4-6:

$\bf Case$  $\bf4:$ The Poisson brackets {\upshape\eqref{eq42}}, {\upshape\eqref{eq52}} allow to describe the dynamics of normal states of magnets with the SU(3) Hamiltonian symmetry. The state of the studied magnets will have the exchange energy density, which is the function of the matrix  $g_{\alpha\beta}({\bf x})$ and its gradient $e({\bf x})=e(\hat{g}({\bf x}), \nabla\hat{g}({\bf x}))$. For the matrix $g_{\alpha\beta}({\bf x})$, the equation {\upshape\eqref{eq10}} is true, the rank of which is equal to three. In terms of real matrices  $q_{\alpha\beta}$ and $\varepsilon_{\alpha\beta}=\varepsilon_{\alpha\beta\gamma}s_{\gamma}/2$, this equation can be rewritten in the form of two matrix equations [10]:
\begin{equation}
\label{eq54}
\begin{split}
&\dot{\hat{q}}({\bf x})=\left[\hat{\varepsilon}({\bf x}), \frac{\delta\hat{H}(\hat{q}, \hat{\varepsilon})}{\delta q({\bf x})}\right]-\left[\hat{q}({\bf x}), \frac{\delta\hat{H}(\hat{q}, \hat{\varepsilon})}{\delta \varepsilon({\bf x})}\right],
\\ &\dot{\hat{\varepsilon}}({\bf x})=\left[\frac{\delta\hat{H}(\hat{q}, \hat{\varepsilon})}{\delta q({\bf x})}, \hat{q}({\bf x})\right]+\left[\frac{\delta\hat{H}(\hat{q}, \hat{\varepsilon})}{\delta \varepsilon({\bf x})}, \hat{\varepsilon}({\bf x})\right].
\end{split}
\end{equation}
Here, $\left(\frac{\delta\hat{H}}{\delta\varepsilon}\right)_{\alpha\beta}=\varepsilon_{\alpha\beta\gamma}\frac{\delta H}{\delta s_{\gamma}}$. Using the SU(3) Hamiltonian symmetry property {\upshape\eqref{eq34}}, we obtain these equations in the form of differential conservation laws
\begin{align*}
\dot{\hat{q}}=-\nabla_k\left[\hat{\varepsilon}, \frac{\partial\hat{e}(\hat{q}, \hat{\varepsilon})}{\partial\nabla_k q}\right]-\nabla_k\left[\frac{\partial\hat{e}(\hat{q}, \hat{\varepsilon})}{\partial\nabla_k\varepsilon}, \hat{q}\right],
\\ \dot{\hat{\varepsilon}}=-\nabla_k\left[\frac{\partial\hat{e}(\hat{q}, \hat{\varepsilon})}{\partial\nabla_kq}, \hat{q}\right]-\nabla_k\left[\frac{\partial\hat{e}(\hat{q}, \hat{\varepsilon})}{\partial\nabla_k\varepsilon}, \hat{\varepsilon}\right].
\end{align*}
The analytical form of the SU (3) symmetric exchange Hamiltonian can be established by analogy with the Heisenberg Hamiltonian. Let us express the homogeneous part of the magnetic energy density in terms of the Casimir invariants  $g_2$ and $g_3$. For simplicity, we will further consider the dependence of the exchange energy only from the invariant  $g_2$
\begin{center}
$H(g_2)=-2\int{d^3xd^3x'J(|{\bf x-x'}|)tr\hat{g}({\bf x})\hat{g}({\bf x'})}$.
\end{center}
Here $J(|{\bf x-x'}|)$ is the exchange integral of two-particle magnetic interaction. The energy density corresponding to the Hamiltonian  $H(g_2)$ have the form
\begin{equation}
\label{eq55}
e({\bf x})=-2Jg_2({\bf x})+\bar{J}tr\nabla\hat{g}({\bf x})\nabla\hat{g}({\bf x}).
\end{equation}
Exchange constants in the energy density {\upshape\eqref{eq55}} are selected so that when there are no quadrupole degrees of freedom, this expression would transfer into the formula {\upshape\eqref{eq44}}. For the energy of the form {\upshape\eqref{eq55}}, the equations {\upshape\eqref{eq54}} take the form
\begin{center}
$\dot{\hat{q}}=\bar{J}[\triangle\hat{\varepsilon}, \hat{q}]+\bar{J}[\triangle\hat{q}, \hat{\varepsilon}], ~~~~ \dot{\hat{\varepsilon}}=\bar{J}[\hat{q}, \triangle\hat{q}]+\bar{J}[\triangle\hat{\varepsilon}, \hat{\varepsilon}]$.
\end{center}

$\bf Case$ $\bf 5$: Spin vectors  $\bf s(x)$ and the tensor order parameter $w_{\alpha\beta}({\bf x})$ will form a closed Poisson bracket subalgebra  {\upshape\eqref{eq42}}, {\upshape\eqref{eq53}}. This is a physically new case of the SO(3) symmetry breaking, which is absent in magnets with the spin s=1/2. Given that   $e=e({\bf s}, \nabla{\bf s}, \hat{w}, \nabla\hat{w})$, we will obtain the dynamic equations
\begin{align*}
\dot{s}_{\alpha}=\varepsilon_{\alpha\beta\gamma}\left(\frac{\delta H}{\delta s_{\beta}}s_{\gamma}+2\frac{\delta H}{\delta w_{\beta\lambda}}w_{\gamma\lambda}\right), ~~~~ \dot{w}_{\beta\gamma}=-(\varepsilon_{\alpha\gamma\rho}w_{\rho\beta}+\varepsilon_{\alpha\beta\rho}w_{\rho\gamma})\frac{\delta H}{\delta s_{\alpha}}.
\end{align*}
The symmetry property {\upshape\eqref{eq33}} for the energy density allows to transform the first equation to the form
\begin{align*}
\dot{s}_{\alpha}=-\nabla_kj_{\alpha k}, ~~~~ j_{\alpha k}=\varepsilon_{\alpha\beta\gamma}\left(\frac{\partial e}{\partial\nabla_ks_{\beta}}s_{\gamma}+2\frac{\partial e}{\partial\nabla_kw_{\beta\lambda}}w_{\gamma\lambda}\right).
\end{align*}
Let us choose the exchange energy density $e=e_o+e_n$ in the form $e_o=Js^2/2+Atr\hat{w}^2/2$, $e_n=Btr(\nabla\hat{w})^2/2$. Non-linear dynamic equations for this type of magnetic ordering will take the form
\begin{center}
$\dot{s}_{\alpha}=-2B\varepsilon_{\alpha\beta\gamma}\triangle w_{\beta\lambda}w_{\gamma\lambda}, ~~~~ \dot{w}_{\beta\gamma}=-Js_{\alpha}(\varepsilon_{\alpha\gamma\rho}w_{\rho\beta}+\varepsilon_{\alpha\beta\rho}w_{\rho\gamma})$.
\end{center}

$\bf Case$ $\bf6$: The set of magnetic dynamic values consist of Hermitian matrices  $a_{\alpha\beta}({\bf x})$ and  $g_{\alpha\beta}({\bf x})$. Formulas {\upshape\eqref{eq42}}, {\upshape\eqref{eq52}}, {\upshape\eqref{eq53}} allow us to formulate the dynamic equations for the magnets with the spin s=1 in the case of the complete SU(3) symmetry breaking. The Hamiltonian is a functional of the matrices  $\hat{g}({\bf x})$,  $\hat{a}({\bf x})$:   $H=H(\hat{g}({\bf x}), \hat{a}({\bf x}))$. The dynamic equations for the $3\times3$  matrices $\hat{a}({\bf x})$  and  $\hat{g}({\bf x})$ will have the form {\upshape\eqref{eq31}}, {\upshape\eqref{eq35}}, {\upshape\eqref{eq37}}. As the symmetry properties {\upshape\eqref{eq34}} for the energy density does not change, we will obtain flux density of magnetic degree of freedom in the considered case
\begin{align*}
\hat{j}_k=i\left[\hat{g}, \frac{\partial\hat{e}}{\partial\nabla_kg}\right]+i\left[\hat{a}, \frac{\partial\hat{e}}{\partial\nabla_ka}\right].
\end{align*}
Let us choose the exchange energy density in the form  $e=e_o+e_n$,  $e_o=Jtr\hat{g}^2/2+Atr\hat{a}^2/2$, $e_n=\bar{J}tr(\nabla\hat{g})^2/2+Btr(\nabla\hat{a})^2/2$. Non-linear dynamic equations will take the form
\begin{center}
$\dot{\hat{g}}=-i(\bar{J}[\hat{g}, \triangle\hat{g}]+B[\hat{a}, \triangle\hat{a}]), ~~~~ \dot{\hat{a}}=i[\hat{a}, J\hat{g}-\bar{J}\triangle\hat{g}]$.
\end{center}
\section{Magnets with the spin s=3/2 }
According to the approach of the section 2, the kinematic part of the Lagrangian of the magnets with spin 3/2 can be represented in the form of {\upshape\eqref{eq5}}, where $\hat{b}$ and $\hat{a}$ are Hermitian  $4\times4$ matrices. Let us introduce the values characterizing the magnetic degrees of freedom. This is a scalar -- $g$, three magnetic vectors -   $\bf u, s, v$ and a quadrupole matrix - $q_{\alpha\beta}$, related to the matrix elements $\hat{g}$ by the formulas
\begin{equation}
\label{eq61}
\begin{split}
&s_{\alpha}=2i\varepsilon_{\alpha\beta\gamma}g_{\beta\gamma}, ~~~~ u_{\alpha}=2i(g_{\alpha4}-g_{4\alpha}), \\g=g_{44}, ~~~~ &v_{\alpha}=(g_{\alpha4}+g_{4\alpha})/2, ~~~~ q_{\alpha\beta}=(g_{\alpha\beta}+g_{\beta\alpha}-2g_{\gamma\gamma}\delta_{\alpha\beta})/2.
\end{split}
\end{equation}
The Greek letters here denote spin indices ranging over values $\alpha,\beta,\gamma,..=1,2,3$. It is easy to see that for the vector   $s_{\alpha}$, the Poisson brackets {\upshape\eqref{eq42}} will be valid in view of {\upshape\eqref{eq61}}. For the vectors $u_{\alpha}$ and $s_{\alpha}$, the Poisson brackets have the form
\begin{equation}
\label{eq62}
\{u_{\alpha}({\bf x}), u_{\beta}({\bf x}')\}=\delta({\bf x-x'})\varepsilon_{\alpha\beta\gamma}s_{\gamma}({\bf x}), ~~~~ \{s_{\alpha}({\bf x}), u_{\beta}({\bf x}')\}=\delta({\bf x-x'})\varepsilon_{\alpha\beta\gamma}u_{\gamma}({\bf x}).
\end{equation}
For the variables  $s_{\alpha}$ and $q_{\alpha\beta}$, the Poisson brackets coincide with the formulas {\upshape\eqref{eq55}}. For the vectors  $s_{\alpha}$ and $v_{\alpha}$, we obtain
\begin{equation}
\label{eq63}
\{s_{\alpha}({\bf x}), v_{\beta}({\bf x}')\}=\delta({\bf x-x'})\varepsilon_{\alpha\beta\gamma}v_{\gamma}({\bf x}), ~~~~ \{v_{\alpha}({\bf x}), v_{\beta}({\bf x}')\}=\delta({\bf x-x'})\varepsilon_{\alpha\beta\gamma}s_{\gamma}({\bf x}).
\end{equation}
For the remaining variables, we find
\begin{equation}
\label{eq64}
\begin{split}
&\{u_{\alpha}({\bf x}), q_{\beta\gamma}({\bf x'})\}=\delta({\bf x-x'})(\delta_{\alpha\beta}v_{\gamma}({\bf x})+\delta_{\alpha\gamma}v_{\beta}({\bf x})-2\delta_{\beta\gamma}v_{\alpha}({\bf x})/3)/2,\\ &\{v_{\alpha}({\bf x}), q_{\beta\gamma}({\bf x'})\}=\delta({\bf x-x'})(\delta_{\alpha\beta}u_{\gamma}({\bf x})+\delta_{\alpha\gamma}u_{\beta}({\bf x})-2\delta_{\beta\gamma}u_{\alpha}({\bf x})/3)/2,\\ &\{u_{\alpha}({\bf x}), g({\bf x}')\}=-\delta({\bf x-x'})v_{\alpha}({\bf x}), ~~~~ \{v_{\alpha}({\bf x}), g({\bf x}')\}=\delta({\bf x-x'})u_{\alpha}({\bf x}),\\ &\{u_{\alpha}({\bf x}), v_{\beta}({\bf x}')\}=2\delta({\bf x-x'})(4\delta_{\alpha\beta}g({\bf x})/3-q_{\alpha\beta}({\bf x})).
\end{split}
\end{equation}
The Poisson bracket algebra contains 15 magnetic degrees of freedom and allow us to describe the dynamics of normal states of magnets with the SU(4) symmetry. Three Casimir invariants reduce the number of independent degrees of freedom to twelve. We do not cite these formulas because of their awkwardness.

Let us consider the Poisson bracket subalgebras using the formulas {\upshape\eqref{eq42}}, {\upshape\eqref{eq52}}, {\upshape\eqref{eq62}}-{\upshape\eqref{eq64}} and distinguish those, which do not coincide with the previously discussed cases of magnetic ordering from sections 2-4. They will include cases 7-9:

$\bf Case$ $\bf 7$. The set of degrees of freedom of the magnet consists of vectors $\bf s, u, (s, v)$ and corresponds to the SU(2)$\times$SU(2) symmetry. Using formulas {\upshape\eqref{eq42}}, {\upshape\eqref{eq62}}, we will obtain the dynamic equations for these spin vectors
\begin{align*}
\dot{u}_{\alpha}=\varepsilon_{\alpha\beta\gamma}\left(\frac{\delta H}{\delta s_{\beta}}u_{\gamma}+\frac{\delta H}{\delta u_{\beta}}s_{\gamma}\right), ~~~~ \dot{s}_{\alpha}=\varepsilon_{\alpha\beta\gamma}\left(\frac{\delta H}{\delta u_{\beta}}u_{\gamma}+\frac{\delta H}{\delta s_{\beta}}s_{\gamma}\right).
\end{align*}
The property of the SU(2)$\times$SU(2) symmetry is equivalent to the formulas
\begin{equation}
\label{eq65}
\{U_{\alpha}, e({\bf x})\}=0, ~~~~ \{S_{\alpha}, e({\bf x})\}=0,
\end{equation}
where the following notations  $U_{\alpha}\equiv\int{d^3xu_{\alpha}({\bf x})}$, $S_{\alpha}\equiv\int{d^3xs_{\alpha}({\bf x})}$ are introduced. Using {\upshape\eqref{eq65}}, we will obtain the dynamic equations for two spin vectors in the form of differential conservation laws
\begin{equation}
\label{eq66}
\begin{split}
&\dot{u}_{\alpha}=-\nabla_kj_{\alpha k}^{(u)}, ~~~~ j_{\alpha k}^{(u)}=\varepsilon_{\alpha\beta\gamma}\left(\frac{\partial e}{\partial\nabla_ks_{\beta}}u_{\gamma}+\frac{\partial e}{\partial\nabla_ku_{\beta}}s_{\gamma}\right), \\
&\dot{s}_{\alpha}=-\nabla_kj_{\alpha k}, ~~~~ j_{\alpha k}=\varepsilon_{\alpha\beta\gamma}\left(\frac{\partial e}{\partial\nabla_ku_{\beta}}u_{\gamma}+\frac{\partial e}{\partial\nabla_ks_{\beta}}s_{\gamma}\right).
\end{split}
\end{equation}
Let us choose the inhomogeneous part of exchange energy density in the form $e_n=\bar{J}tr(\nabla\hat{g}^{(a)})^2/2=\bar{J}\left[(\nabla_k{\bf s})^2+(\nabla_{k}{\bf u})^2\right]/2$. Therefore, we can rewrite the equations of {\upshape\eqref{eq66}} as
\begin{align*}
\dot{u}_{\alpha}=-\bar{J}\varepsilon_{\alpha\beta\gamma}(\triangle s_{\beta}u_{\gamma}+\triangle u_{\beta}s_{\gamma}), ~~~~ \dot{s}_{\alpha}=-\bar{J}\varepsilon_{\alpha\beta\gamma}(\triangle u_{\beta}u_{\gamma}+\triangle s_{\beta}s_{\gamma}).
\end{align*}

$\bf Case$ $\bf 8$. We introduce the polarization density matrix by relation   $\hat{\rho}({\bf x})=\hat{I}/4+\hat{g}({\bf x})$. The expand of the physical matrix  $\hat{g}({\bf x})$ will have the form
\begin{equation}
\label{eq67}
\hat{g}({\bf x})=\gamma_5({\bf x})\hat{\gamma}_5+\gamma_{\mu}({\bf x})\hat{\gamma}_{\mu}+\bar{\gamma}_{\mu}({\bf x})\hat{\bar{\gamma}}_{\mu}+\sigma_{\mu\nu}({\bf x})\hat{\sigma}_{\mu\nu}/2.
\end{equation}
The set of irreducible matrices (4$\times$4)  $\hat{\gamma}_{a}$, ($a=1,..,15$) consists of $\hat{\gamma}_{\mu}$ Dirac matrices and matrices of the form $\hat{\bar{\gamma}}_{\mu}\equiv i\hat{\gamma}_5\hat{\gamma}_{\mu}$,  $\hat{\gamma}_5\equiv i\hat{\gamma}_1\hat{\gamma}_2\hat{\gamma}_3\hat{\gamma}_4$, $\hat{\sigma}_{\mu\nu}\equiv i\left[\gamma_{\mu}, \hat{\gamma}_{\nu}\right]/2$. By virtue of tracelessness condition   $tr\hat{\gamma}_{a}=0$ and orthogonality condition $tr\hat{\gamma}_a\hat{\gamma}_b=4\delta_{ab}$ of basic matrices, the values  $\gamma_a\equiv\gamma_5({\bf x}), \gamma_{\mu}({\bf x}), \bar{\gamma}_{\mu}({\bf x}), \sigma_{\mu\nu}({\bf x})$ in {\upshape\eqref{eq67}} connected with the matrix  $\hat{g}({\bf x})$ by relation  $\gamma_a({\bf x})=tr\hat{g}({\bf x})\hat{\gamma}_a/4$. It is well known that the algebra Dirac matrices  $\hat{\gamma}_a$ is six-dimensional. Independent matrices are infinitesimal operators $\hat{S}_{ik} (\hat{S}_{ik}=-\hat{S}_{ki}), (i,k=1,..,6)$ of orthogonal group in the six-dimensional space [26,27], that is  $2\hat{S}_{6\mu}=\hat{\gamma}_{\mu}$, $2\hat{S}_{65}=\hat{\gamma}_5$, $2\hat{S}_{\mu5}=\hat{\gamma}_{\mu}$, $2\hat{S}_{\mu\nu}=\hat{\sigma}_{\mu\nu}$. Using these relations and the formula [28]
\begin{center}
$i\left[\hat{S}_{ik}, \hat{S}_{lm}\right]=\delta_{im}\hat{S}_{kl}-\delta_{il}\hat{S}_{km}-\delta_{km}\hat{S}_{il}+\delta_{kl}\hat{S}_{im}$
\end{center}
it is easy to obtain non-trivial Poisson brackets for values  $\gamma_{a}({\bf x})$:
\begin{equation}
\label{eq68}
\begin{split}
&\{\gamma_{5}({\bf x}), \gamma_{\mu}({\bf x}')\}=-\delta({\bf x-x'})\bar{\gamma}_{\mu}({\bf x})/2, ~~~~ \{\gamma_5({\bf x}), \bar{\gamma}_{\mu}({\bf x'})\}=\delta({\bf x-x'})\gamma_{\mu}({\bf x})/2, \\
&\{\gamma_{\mu}({\bf x}), \gamma_{\nu}({\bf x}')\}=-\delta({\bf x-x'})\sigma_{\mu\nu}({\bf x})/2, ~~~~ \{\bar{\gamma}_{\mu}({\bf x}), \bar{\gamma}_{\nu}({\bf x}')\}=-\delta({\bf x-x'})\sigma_{\mu\nu}({\bf x})/2, \\
&\{\gamma_{\mu}({\bf x}), \bar{\gamma}_{\nu}({\bf x}')\}=-\delta({\bf x-x'})\delta_{\mu\nu}\gamma_5({\bf x})/2, \\
&\{\sigma_{\mu\nu}({\bf x}), \sigma_{\lambda\rho}({\bf x'})\}=-\delta({\bf x-x'})(\delta_{\nu\lambda}\sigma_{\mu\rho}({\bf x})+\delta_{\nu\rho}\delta_{\lambda\mu}({\bf x})+\delta_{\mu\rho}\sigma_{\nu\lambda}({\bf x})+\delta_{\mu\lambda}\sigma_{\rho\nu}({\bf x}))/2, \\
&\{\gamma_{\lambda}({\bf x}), \sigma_{\mu\nu}({\bf x}')\}=\delta({\bf x-x'})(\gamma_{\nu}({\bf x})\delta_{\lambda\mu}-\gamma_{\mu}({\bf x})\delta_{\lambda\nu})/2, \\
&\{\bar{\gamma}_{\lambda}({\bf x}), \sigma_{\mu\nu}({\bf x}')\}=\delta({\bf x-x'})(\bar{\gamma}_{\nu}({\bf x})\delta_{\lambda\mu}-\bar{\gamma}_{\mu}({\bf x})\delta_{\lambda\nu})/2.
\end{split}
\end{equation}

The inhomogeneous part of the exchange energy will have the form {\upshape\eqref{eq38}}. Using the symmetry properties  $\{\Gamma_a, e({\bf x})\}=0$, where  $\Gamma_a=\int{d^3x\gamma_a({\bf x})}$, and considering {\upshape\eqref{eq68}}, we will obtain the dynamic equations

\begin{center}
$\dot{\gamma}_5=2\bar{J}(\bar{\gamma}_{\mu}\triangle\gamma_{\mu}-\gamma_{\mu}\triangle\bar{\gamma}_{\mu}), ~~~~ \dot{\gamma}_{\mu}=2\bar{J}(\gamma_5\triangle\bar{\gamma}_{\mu}-\bar{\gamma}_{\mu}\triangle\gamma_5+\sigma_{\mu\nu}\triangle\gamma_{\nu}-\gamma_{\nu}\triangle\sigma_{\mu\nu})$, \\ $\dot{\bar{\gamma}}_{\mu}=2\bar{J}(\gamma_{\mu}\triangle\gamma_5-\gamma_5\triangle\gamma_{\mu}+\bar{\gamma}_{\nu}\triangle\sigma_{\nu\mu}-\sigma_{\nu\mu}\triangle\bar{\gamma}_{\nu}),$ \\ $\dot{\sigma}_{\mu\nu}=2\bar{J}(\gamma_{\nu}\triangle\gamma_{\mu}-\gamma_{\mu}\triangle\gamma_{\nu}+\bar{\gamma}_{\nu}\triangle\bar{\gamma}_{\mu}-\bar{\gamma}_{\mu}\triangle\bar{\gamma}_{\nu}+\sigma_{\mu\lambda}\triangle\sigma_{\lambda\nu}-\sigma_{\lambda\nu}\triangle\sigma_{\mu\lambda})$.
\end{center}

If the additive integrals of motion of magnets are the only anti-symmetric matrix  $\Sigma_{\mu\nu}$,  $\{\Sigma_{\mu\nu}, H\}=0$, then using the functional hypothesis of reduced description for the densities of these values  $H=H(\hat{\sigma}, \nabla\hat{\sigma})$, we get SO(4) symmetric equations of dynamics
\begin{center}
$\dot{\sigma}_{\mu\nu}=2\bar{J}(\sigma_{\mu\lambda}\triangle\sigma_{\lambda\nu}-\sigma_{\lambda\nu}\triangle\sigma_{\mu\lambda})$.
\end{center}
Here inhomogeneous exchange energy density has the form  $e_{n}({\bf x})=\bar{J}(\nabla_k\sigma_{\mu\nu}({\bf x}))^2/4.$

$\bf Case$ $\bf9$.  The decomposition of the physical matrix $\hat{g}$ will have the form
\begin{center}
$\hat{g}({\bf x})=\gamma_a({\bf x})\hat{\gamma}_a+\gamma_{ab}({\bf x})\hat{\gamma}_{ab}/2$.
\end{center}
The basic set of matrices will contain five matrices  $\hat{\gamma}_a\equiv(\hat{\gamma}_{\mu}, \hat{\gamma}_5)$  and $\hat{\gamma}_{ab}\equiv i\left[\hat{\gamma}_a,\hat{\gamma}_b\right]/2, a<b$, $(a,b=1,2,..,5)$. The set of degrees of freedom of the magnet will consist of values  $\gamma_a({\bf x}), \gamma_{ab}({\bf x})$. Given the algebraic relations
\begin{center}
$i\left[\hat{\gamma}_{ab}, \hat{\gamma}_{c}\right]=2(\delta_{ac}\hat{\gamma}_{b}-\delta_{bc}\hat{\gamma}_{a}), ~~~~ i\left[\hat{\gamma}_{ab}, \hat{\gamma}_{cd}\right]=2(\delta_{ac}\hat{\gamma}_{bd}-\delta_{bc}\hat{\gamma}_{ad}+\delta_{ad}\hat{\gamma}_{cb}-\delta_{bd}\hat{\gamma}_{ca})$,
\end{center}
and the orthogonality conditions
\begin{center}
$tr\hat{\gamma}_{a}=0, ~~~~ tr\hat{\gamma}_{ab}=0, ~~~~ tr\hat{\gamma}_{a}\hat{\gamma}_{b}=4\delta_{ab}, ~~~~ tr\hat{\gamma}_{ab}\hat{\gamma}_{cd}=4(\delta_{ac}\delta_{db}-\delta_{cb}\delta_{ad}), ~~~~ tr\hat{\gamma}_{a}\hat{\gamma}_{bc}=0$,
\end{center}
for values  $\gamma_a({\bf x})=tr\hat{g}({\bf x})\hat{\gamma}_a/4$ and  $\gamma_{ab}({\bf x})=tr\hat{g}({\bf x})\hat{\gamma}_{ab}/4$, we will obtain the Poisson brackets
\begin{center}
$\{\gamma_{a}({\bf x}), \gamma_{b}({\bf x}')\}=-\delta({\bf x-x'})\gamma_{ab}({\bf x})/2$, ~~~~ $\{\gamma_{c}({\bf x}), \gamma_{ab}({\bf x}')\}=\delta({\bf x-x'})(\delta_{ac}\gamma_b({\bf x})-\delta_{bc}\gamma_a({\bf x}))/2$,\\
$\{\gamma_{ab}({\bf x}), \gamma_{cd}({\bf x}')\}=\delta({\bf x-x'})(\delta_{bc}\gamma_{ad}({\bf x})-\delta_{ac}\gamma_{bd}({\bf x})+\delta_{bd}\gamma_{ca}({\bf x})-\delta_{ad}\gamma_{cb}({\bf x}))/2$.
\end{center}
Using the symmetry conditions  $\{\Gamma_a, e({\bf x})\}=0$,  $\{\Gamma_{ab}, e({\bf x})\}=0$, where  $\Gamma_a=\int{d^3x\gamma_a({\bf x})}$,  $\Gamma_{ab}=\int{d^3x\gamma_{ab}({\bf x})}$, we will get the dynamic equations
\begin{center}
$\dot{\gamma}_a=2\bar{J}(\gamma_{ab}\triangle\gamma_b-\gamma_b\triangle\gamma_{ab}), ~~~~ \dot{\gamma}_{ab}=2\bar{J}(\gamma_b\triangle\gamma_a-\gamma_a\triangle\gamma_b+\gamma_{cb}\triangle\gamma_{ac}-\gamma_{ac}\triangle\gamma_{cb})$.
\end{center}
Here, the inhomogeneous magnetic energy density will have the form  {\upshape\eqref{eq38}}.

A special case of SO(5) symmetry corresponds to the presence only integrals of motion  $\Gamma_{ab}$,  $\{\Gamma_{ab}, H\}=0$. The dynamic equations of the density of these integrals of motion, taking the form of the inhomogeneous exchange energy  $e_n=\bar{J}(\nabla_k\gamma_{ab})^2$, have the form
\begin{center}
$\dot{\gamma}_{ab}=2\bar{J}(\gamma_{cb}\triangle\gamma_{ac}-\gamma_{ac}\triangle\gamma_{cb})$.
\end{center}
\section{Conclusions}
The class of non-linear dynamic equations for high-spin magnets, which generalized the Landau-Lifshitz equation, were obtained. The presence of several types symmetry of exchange Hamiltonians and set of Casimir invariants lead to new possible magnetic states. The Hamiltonian approach we have developed allows us to generalize the dynamic theory of magnets for an arbitrary spin case considering the SU(2s+1) symmetry. Non-linear solutions for this class of dynamic equations require their analysis and study.

\end{document}